# Fano resonance assisting plasmonic circular dichroism from nanorice heterodimers for extrinsic chirality

Li Hu[1,3], Yingzhou Huang[1,*], Liang Fang[1], Guo Chen[1], Hua Wei[1], Yurui Fang[2,*]

[1] Soft Matter and Interdisciplinary Research Center, College of Physics, Chongqing University, Chongqing, 400044, P. R. China

[2] Department of Applied Physics, Chalmers University of Technology, Göteborg, SE 412 96, Sweden

[3] School of Computer Science and Information Engineering, Chongqing Technology and Business University, Chongqing, 400067, China

Corresponding Authors' Emails: yzhuang@cqu.edu.cn (Y. Huang), yurui.fang@chalmers.se (Y. Fang)

**In this work, the circular dichroisms (CD) of nanorice heterodimers consisting of two parallel arranged nanorices with the same size but different materials are investigated theoretically. Symmetry-breaking is introduced by using different materials and oblique incidence to achieve strong CD at the vicinity of Fano resonance peaks. We demonstrate that all Au-Ag heterodimers exhibit multipolar Fano resonances and strong CD effect. A simple quantitative analysis shows that the structure with larger Fano asymmetry factor has stronger CD. The intensity and peak positions of the CD effect can be flexibly tuned in a large range by changing particle size, shape, the inter-particle distance and surroundings. Furthermore, CD spectra exhibit high sensitivity to ambient medium in visible and near infrared regions. Our results here are beneficial for the design and application of high sensitive CD sensors and other related fields.**

O‌ptical activity (OA), whose origin can be reduced to a different response of a system to right- and left-circularly polarized light, is a fantastic optical phenomenon discovered more than 100 years ago. It has two manifestation forms, which are optical rotation and circular dichroism (CD)[1]. Based on OA, optical rotator dispersion (ORD) spectroscopy, CD spectroscopy and Raman optical activity (ROA) spectroscopy have been developed to significant analysis methods in the study of medicine diagnosis, crystallography, analytical chemistry, molecular biology and life form in universe[2-7]. However, most nature molecules only manifest very weak optical activity, which greatly limits their applications.

In recent years, chiroplasmonics is a hotspot of current research in plasmonics due to the giant OA in chiral metallic nanostructures which has potential applications in ultra-sensitive sensing. The strong coupling between light and surface plasmons (SPs), which are collective oscillations of free electrons in the interface of metal-dielectric, is responsible for the giant OA. Since the status of SPs in metallic nanostructures is sensitive to the shape, size, material and configuration of structures, it offers a flexible way to tune OA effect in a broad band from ultraviolet to near-infrared. So far, giant OA due to different mechanisms have been intensively studied in various chiral plasmonic nanostructures, such as chiral metal particles[8-10], pairs of mutually twisted planar metal patterns[11], single-layered metal saw-tooth gratings[12], planar chiral metal patterns[13-15], DNA based self-assembled metal particles[16,17], helical metal wires[18], etc.

In addition to the above listed intrinsic chiral nanostructures, which are chiral in the sense that they cannot be superimposed on their mirror image by using spatial operation (rotation, translation, etc.), two dimensional (2D) achiral nanostructures can also show giant optical activity under certain conditions（e.g. simply by titling the structures symmetry axis out of incident plane to yield the three dimensional asymmetry）, which is called "extrinsic chirality"[19-21], Extrinsic chirality provides more flexible way to overcome the difficulty in fabrication progress of complex chiral structure and shows even stronger CD than the intrinsic ones. It was firstly observed in metallic nanostructures by N. I. Zheludev in 2008, where extrinsic chirality leads to exceptionally large CD in the microwave region[22]. Following that, extrinsic chirality induced CD was expanded to visible and near-infrared range[23,24]. Very recently, Lu et al. experimentally measured strong CD from single plasmonic nanostructure with extrinsic chirality[25]. Kato et al. observed giant CD from individual carbon nanotubes induced by extrinsic chirality[26]. This means that extrinsic chirality induced CD may be strong enough to detect material in single molecule level. In the same year of 2014, Tian et al. theoretically predicted that Fano resonances have assisting effect on extrinsic CD in 2D nanostructures[27]. A near 100% difference of response to right- and left-circularly polarized light was indicated in higher-order plasmonic resonant modes.

In this work, we further investigate Fano assisting plasmonic CD in heterodimers consisting of two parallel arranged nanorices with the same size but different materials through finite element method (FEM). The influences of materials, geometric parameters, surrounding dielectrics on Fano assisting CD were investigated, which was analyzed by surface charge density and a chiral molecule model. The results indicate that the Au-Ag nanorice heterodimer showed good chiral optical response, which may have promising applications in sensitive sensing.

## Result

**Fano assisting CD illustration.** To address those concerns and illustrate the principle of the Fano assisting CD, we first focus on the connection between Fano resonance of plasmon and the mechanism of CD. The CD mechanism of extrinsically chiral structures is not as obvious as the intrinsic ones. For the intrinsically chiral structures, the response of structures can be easily found out by looking at the chiral structure and the opposite rotation of left/right polarized light (LCP/RCP). While for extrinsically chiral structures, analogical to the CD of molecules, we can get the response difference from initial interactions of electric and magnetic dipoles ($A^{\pm} = \frac{\omega}{2}(\alpha''|\vec{E}|^2 + \gamma''|\vec{B}|^2) \pm G''\omega \text{Im}(\vec{E}^* \cdot \vec{B})$, where $G''$ is the imaginary part of the isotropic mixed electric-magnetic dipole polarizability)[28]. That means the induced electric and magnetic dipoles must have nonzero component in the plane normal to the incident light (Fig. 1a). To obtain plasmonic nanostructures of extrinsic chirality, several criteria should be met with which are: the structure has no inversion center, no reflection symmetry in the normal plane of light propagation, no inversion or mirror rotation axis along light propagation. Off-normal incident is an easy way to realize this. Plasmonic Fano resonance is induced by the interference between the narrow subradiant and broad super radiant plasmonic modes[29-31], where the subradiant mode consists of two opposite electric dipoles. The opposite dipoles are equivalent to a magnetic dipole and an additional electric dipole if the two dipoles are not the same momentum (Fig. 1a). Where the Fano resnonance happens, the subradiant mode is in resonance, which means that the two dipoles have strong magnet like a metal magnetron. Thus the Fano effect will have an enhancement for the CD. If the multiple resonance modes are excited, more magnetrons will appear with different momentum direction, and superimpose coherently because of the phase difference of them.

**Simulation model.** Figure 1 b-e show the simulation realization and basic results of this paper. The simulations were performed with nanorice[32,33] dimer structures using FEM (Comsol Multiphysics 4.3a). The sketch of the simulation model in this work is illustrated in Figure 1b & c. The heterodimer consists of two nanorices with the same size but different materials. Here the heterodimer locates in x-y plane and the incident light with LCP or RCP propagated in x-z plane with θ = 45˚ off the z axis. The long axis of the nanorice was set as $l$, the short one was set as $d$, the refractive index of surrounding medium was set as *n*. The permittivities of materials were adopted according to Johnson and Christy's works[34]. The intensity of CD was defined as "$CD = \sigma_L - \sigma_R$", in which $\sigma_L$ and $\sigma_R$ were extinction cross sections under LCP and RCP excitation respectively. Sphere structures calculated with Mie code and nanorice dimer structures calculated with FDTD were performed to check the validity of the COMSOL simulations. Almost the same spectra were gotten.

**The influence of materials on plasmonic CD.** Considering the two nanorices in each dimer were of the same size here, the materials of the two nanorices should be different, thus the momentums of the two rices are different, to satisfy the extrinsically chiral excitation criteria. Therefore, various groups of nanorice materials were investigated firstly. Two nanorices (*l=240 nm, d=60 nm*) were parallel placed to each other with a 10 nm gap in a medium with refractive index n = 1.1. Illuminated by LCP and RCP light respectively, the heterodimers composed of various materials (Au-Au, Au-Ag，Au-TiO$_2$, Au-SiO$_2$, Au-Al$_2$O$_3$) exhibited different CD phenomena. As Figure 1d shown, the heterodimer has no optical activity ($CD = 0$) when the two nanorices composed of the same materials (e.g. Au-Au). However, the other heterodimers which composed of two different materials show strong CD effect. To address this influence of materials on plasmonic CD, surface charge density distributions were analyzed in Figure 1e. CD effect in heterodimer could be understood through the interaction of the electric and magnetic responses[28]. The induced electric dipole $\vec{d}$ represented by red line arrows and the magnetic dipole $\vec{m}$ represented by black circle arrows could be analyzed clearly through surface charge distributions in Figure 1e on heterodimers with RCP light illuminating. In the Au-Au dimer, only the electric dipole is induced that it could not produce the CD effect, as the exactly same momentum of the two rices makes the subradiant mode totally dark so as that the light cannot excite this mode. While in other dimers (e.g. Au-Ag, Au-TiO$_2$), both the electric and magnetic dipoles are generated that their interaction are responsible for the strong CD effects. One should pay attention to that the induced electric and magnetic dipoles are not enough for the generation of CD effects. The oblique illuminating is also necessary that there should be projections of electric and magnetic dipoles in the plane perpendicular to the incident wave vector to yield the nonzero 'chiral response'. The stronger CD effect in whole plasmonic metal heterodimer in our work was attributed to the much larger free charge density in metal (thus larger momentum and interaction) compared to dielectrics. Therefore, the Au-Ag heterodimer was focused on in the following studies.

**The analysis of hybrid modes and Fano resonance.** To obtain further understanding of Fano assisting CD phenomenon, the plasmonic Fano resonance in nanorice heterodimer was investigated. Here linearly polarized light in x-z plane instead of circular polarization light was adopted in model shown in Figure 2. Under the oblique illuminating, there are two peaks (not only common dipole mode but also quadruple mode) in the extinction spectrum of individual Au or Ag nanorice (blue or red curve in Figure 2a) because of phase delay along the rice. The extinction spectrum of heterodimer hybridized from individual rice shows four resonant peaks (Figure 2b), which are asymmetric. The surface charge density distributions shown in Figure 2c illustrate the resonant modes of the four peaks for the heterodimer. Together with the hybrid energy diagram in Figure 2d, one could see that peak 1 is the bonding mode of dipole mode of Ag and Au nanorices. Peak 2 is a bonding mode of the Ag dipole and Au quadruple mode, where the quadruple mode is induced by the charge on the other rice. This mode is more like an antibonding mode of Ag dipole and Au dipole modes, but the induced small quadruple lower down the energy. Peak 3 is a bonding mode of Ag $l=2$ and Au $l=2$ modes. Peak 4 is an antibonding mode of Ag $l=2$ mode and Au $l=2$ mode. From the surface charge density distributions we can see that peak 1 and peak 3 are subradiant modes, while peak 2 and 4 are superradiant modes. The overlap and interferences of subradiant modes and superradiant modes form Fano resonance profile. As shown in Fig.1, all of the subradiant modes in such a configuration are equivalent to a magneton and an electric dipole, which is necessary for the plasmonic CD. The nano metallic magneton thus will cause both Fano and CD effect. When the Fano is in strong resonance, the system will have a very strong magnetic dipole momentum, which causes a big CD value ( $\propto (\text{Im}(\vec{d} \cdot \vec{m}) * \text{Im}(\vec{E}^* \cdot \vec{B}))$ ). In such configuration, total dark mode will cause strongest magnetic dipole momentum, but it is hard to interact with electric dipole. In this point of view, a Π configuration could make stronger enhancement for CD.

**Optical activity with different gaps between the two nanorices.** From the hybrid diagram we know that the Fano resonance have a strong dependence on the gap of heterodimer. Bigger gap will reduce the interaction of the two basic modes and thus the splitting of the two peak (peak 1 and 2, peak 3 and 4) will becomes smaller. Meanwhile bigger gap will cause weaker magnetic dipole momentum, so the CD will become smaller as well. Figure 3 shows the CD dependence on the gap. When the gap increased from 10 nm to 50 nm, the extinction cross sections change little but the CD spectra change more, which is a demonstration of the point. To quantitatively show the Fano-assisted CD, we choose the most obvious asymmetric peak (peak 4) to do the Fano fitting with $I = A\frac{(q\gamma + \omega - \omega_0)^2}{(\omega - \omega_0)^2 + \gamma^2} + B$ (where q is the Fano asymmetry factor, $\gamma$ is the width of the Fano resonance peak, $\omega_0$ is the resonance position, A is the resonance amplitude and B is the background) for RCP extinction spectra (inset in Figure 3a). The Fano asymmetry factors are gotten as q = -0.588, -0.410 and -0.288 for g = 10 nm, 20 nm and 50 nm respectively. One can clearly see that the structure with larger Fano asymmetry factor (absolute value) in spectrum has larger CD value.

**The effect of the structure factors on plasmonic CD.** Since the structure factors (e.g. size and shape) play a great impact on the surface plasmon in metal nanostructures, their influence on Fano resonance and CD effects were investigated then, which was shown in Figure 4. To simplify the model, the heterodimers with various sizes but the same aspect ratio ($l/d$= 4) were studied first in this section. It can be seen that for individual Ag nanorice (Fig. 4a), with increasing of the size the intensity of the resonant modes especially the quadruple mode increases markedly, the resonant wavelength have a significant red-shift as expected, and the distance increases between the dipole and quadruple modes. When the material of the nanorice changed to Au, the extinction spectra show similar property (supporting information, Fig S2). It can be deduced from the above analysis that if the sizes of Au and Ag nanorices increase at the same time the hybridized modes would red-shift, the extinction cross sections of higher-order modes would increase more relatively and the asymmetric Fano profile would become more distinctive. The deduction was well verified in Figure 4b, which shows the extinction spectra of the Au-Ag heterodimer excited by LCP and RCP light. Besides the above analysis, we can see from Figure 4b that there are remarkable differential of the optical activities to LCP and RCP, especially in the vicinity of the four Fano resonance peaks. Thus Figure 4c shows that with the size increasing the intensity of CD signals increase dramatically, higher-order modes

especially，and all the peaks position red-shift. Therefore the Fano resonance has an "assisting effect" on CD signals.

To investigate the effect from the shape of nanorice on the localized surface plasmon resonance, Fano interference and the CD effect, we tuned the aspect ratio (l/d = 3, 4, 6) of the nanorices by changing the short axis with fixed longer axis (*l=240 nm*), as shown in Figure 4(d-f). Figure 4d shows the extinction spectra of single Ag nanorices with different aspect ratio. The dipole modes and quadruple modes blue-shift and the intensity of dipole modes decrease but the quadruple mode increase with increasing the short axis of the nanorice. At the same time, the gap decreases between the dipole and quadruple modes. Correspondingly, when the aspect ratio of the two nanorices decrease from 6 to 3, all the plasmonic modes of the heterodimer which excited by LCP and RCP light blue-shift, and the Fano profile of the extinction spectra become more distinctive due to the inducing energy overlap of the long-wavelength modes and the short-wavelength modes, as shown in Figure 4e. Figure 4f shows with decreasing the aspect ratio that the CD response become stronger because of the increase of the Fano interference. To quantitatively show the Fano-assisted CD like in Figure 3, we choose the most obvious asymmetric peak (peak 4) as well to do the Fano fitting for RCP extinction spectra (inset in Figure 4e as they have similar size (l = 240 nm), which makes relation between the asymmetric factor and CD comparable. The Fano asymmetry factors are gotten as q = -0.889, -0.588 and -0.309 for d = 80 nm, 60 nm and 40 nm respectively. This once more confirms that the structure with larger Fano asymmetry factor (absolute value) in spectrum has larger CD value, and they are comparable with the relation in Figure 3.

**Sensitivity of the Fano assisting CD to the surroundings.** One of the main arguments is to investigating plasmonic chiral structures sensing. In the recent context, experimental and theoretical studies show that the plasmonic structures can enhance the chiral signals dramatically because of the strongly enahanced superchiral near field. Han. Bing[35] also shows that the plasmonic chiral structures have the selective enhancement for the chiral molecules. Plasmonic Fano resonance has wide applications for sensing as well because the big figure of merit (*FOM*). Based on the above analysis, it is very clear that the Fano assisting plasmonic response for LCP and RCP is very big, even close to 100%. Therefore, the Fano assisting CD of plasmonic structures is a very good candidate for chiral molecules sensing. The different response (absorption) of the same molecules with left or right handedness will cause different permittivities. So the CD response can be used for sensing and distinguishing the handedness of the target molecules. And the sensing is not limited only for chiral molecules but also available for normal sensing. To show that CD response contributes to the sensitivity of the plasmon resonance, different media with the refractive index varied from 1.1 to 1.4 are considered around the heterodimer. Figure 5a shows the CD signal of the Au-Ag heterodimer (*l=240nm, d=60 nm, g=10 nm*) embedded in different matrices. As expected, all modes exhibit significant red-shift with increasing the refraction index of the embedding medium. Figure 5b shows that the resonance peaks shift in wavelength with respect to the refractive index. To estimate the sensitivity of the proposed plasmonic nanostructure, the *FOM* is defined as $FOM = \dfrac{d\lambda_p}{dn \times \Delta\lambda}$ in which $\lambda_p$ is the CD peak position, *n* is the refractive index of media, $\Delta\lambda$ is the full width half maximum. We calculated the *FOM* of the three modes in Figure 5a, for mode 1 the *FOM* from 6.1 to 6.8, the *FOM* from 11.9 to 12.4 of the mode 2, while the *FOM* of mode 3 from 20.7 to 21.6, which are more sensitive than many other nanostructures. A comparison for the *FOM* of Fano assisting CD and only plasmonic peak shift (supporting information, Fig S3 and Fig S4) shows a larger shift for CD than for most of Fano peaks sensing because the dramatic phase change cause narrower peaks. We can see that the CD sensing has around 5 times of the *FOM* larger than Fano sensing for the dipole hybrid mode 3, and the other modes keep similar sensitivity compared with the Fano extinction profile. Such environmental sensitivity of the CD spectra of the nanorice heterodimer holds great potential for monitoring local environmental changes in visible and near infrared region during chemical and biological processes.

**Conclusion and Discussion**

In summary, the subradiant mode of the heterodimer behaves like a magneton. Excitation of the plasmonic magneton will cause a CD effect for the nanorice heterodimer with different materials but the same size. The result shows that the heterodimer composed of Au and Ag nanorices have Fano resonance and strong CD effect. The Fano resonance has an enhancement effect for the CD. A simple quantitative analysis shows that the structure with larger Fano asymmetry factor has stronger CD. When the gap, size and the shape of nanorice are changed, the Fano spectra profile and CD signal change correspondingly, which shows that the Fano resonance has an assisting effect for CD of the nanometal molecules. Excited high order modes show near 100% response difference. In addition, the CD spectra either in visible or near infrared region are highly sensitive to environment, which will have promising applications in both traditional sensing fields and chiral molecules distinguish. Formation of such heterodimer is quite easy in the lab with spin-coating or self-assembly methods[33, 36], and dimers with other shapes have the similar effect as well. So it is very useful in experimental examination.

**Acknowledgements**

We are grateful to Dr. Xiaorui Tian for useful discussions. This work was supported by the National Natural Science Foundation of China (Grant Nos. 11204390, 11004257), Natural Science Foundation Project of CQ CSTC (2014jcyjA40002, 2011jjA90017), Fundamental Research Funds for the Central Universities 106112015CDJZR300003 and Special Fund for Agro-scientific Research in the Public Interest (NO. 201303045).

**Author contributions**

Y. H. and Y. F. supervised the project. L. H. and G. C. did simulation. L. H., Y. H., Y. F., L. F. and H. W. analysed the data. Y. H. and Y. F. wrote the paper.

**Competing financial interests:** The authors declare no competing financial interests.

**Figure Captions**

**Figure 1 | Fano assisting CD illustration, simulation model and The influence of materials on plasmonic CD.** (a) Scheme illustrating the mechanism of Fano assisting CD with electromagnetic responses. Two plasmonic electric dipoles with different momentum hybrid together. The bonding mode is equivlent to a magnetic dipole $\vec{m}$ and a smaller momentum electric dipole $\vec{d}$. A tilted incident light will couple the electric and magnetic dipoles together to have a CD effect. The overlap of antibonding mode and bonding mode of the two electric dipoles will cause a Fano resnonce. When the Fano resonance happens, the bonding mode becomes strongest, so cause a strongest magnetic dipole momentum, which will cause a strong CD profile. (b-c) Schematic representation of an nanorice heterdimer labeled with structure parameters. (b) Top view: The nanorice heterodimer located in the x-y plane, with its normal direction along z axis. (c) Side view: CPL illuminated the heterodimer with angle θ. (d) CD spectra of the heterodimer with different materials. (e) Surface charge density distributions for the three modes of heterodimer with different materials excited by RCP light and marked by colored dots in (d). CPL: circularly polarized light.

**Figure 2 | The analysis of hybrid modes and Fano resonance.** Extinction spectra of the heterodimer (*l=240 nm, d=60 nm*) with g=20 nm under oblique incidence ($\theta = \pi/4$) with linearly polarized light. (a) Extinction spectra of individual nanorices (blue, Au nanorice; red, Ag nanorice) and the surface charge density distributions for dipole and quadrupole modes of Ag nanorice. (b) Extinction spectrum of Au-Ag nanorice heterodimer. (c) Surface charge density distributions for hybridized plasmon modes (marked by numeral in (b)) (top: Au, bottom: Ag). (d) Energy-level diagram describing the plasmon hybridization of the Au-Ag heterodimer.

**Figure 3 | Optical activity with different gaps between the two nanorices.** (a) Extinction spectra of the Au-Ag dimer (*l=240 nm, d=60 nm*) with g=10 nm (blue), 20 nm (red) and 50 nm (black) under LCP and RCP excitation. The inset shows the Fano fitting (green line, $I = A\frac{(q\gamma + \omega - \omega_0)^2}{(\omega - \omega_0)^2 + \gamma^2} + B$) of the indicated peaks for different gaps excited with RCP. The q values reflect the asymmetry of the peaks. (b) The CD spectra of the Au-Ag heterodimer with different gaps.

**Figure 4 | The effect of the structure factors on plasmonic CD.** (a-c) The optical spectra of single Ag nanorice and Au-Ag heterodimer with various size but the same aspect ratio *(l/d=4, g=10 nm)*. (a) The extinction spectra of individual Ag nanorice excited by linear polarized light. (b) The extinction spectra of Au-Ag heterodimer excited by LCP and RCP light. (c) The CD spectra for Au-Ag heterodimer. (d-f) The optical spectra of single Ag nanorice and Au-Ag heterodimer with different aspect ratio *(l/d=3,4,6, g=10 nm)*. (d) The extinction spectra of single Ag nanorice excited by linear polarized light. (e) The extinction spectra of Au-Ag heterodimer excited by LCP and RCP light. The inset shows the Fano fitting (green line, $I = A\frac{(q\gamma + \omega - \omega_0)^2}{(\omega - \omega_0)^2 + \gamma^2} + B$) of the indicated peaks for different gaps excited with RCP. The q values reflect the asymmetry of the peaks. (f) The CD spectra for Au-Ag heterodimer .set a heterodimer as (*l(d)*) in figure.

**Figure 5 | Sensitivity of the Fano assisting CD to the surroundings.** (a) CD spectra of Au-Ag heterodimer (*l=240 nm, d=60 nm, g=10 nm*) in different surroundings. (b) Linear plot of CD peak shifts of different order modes as a function of refractive index n. *FOM* of each mode is calculated.

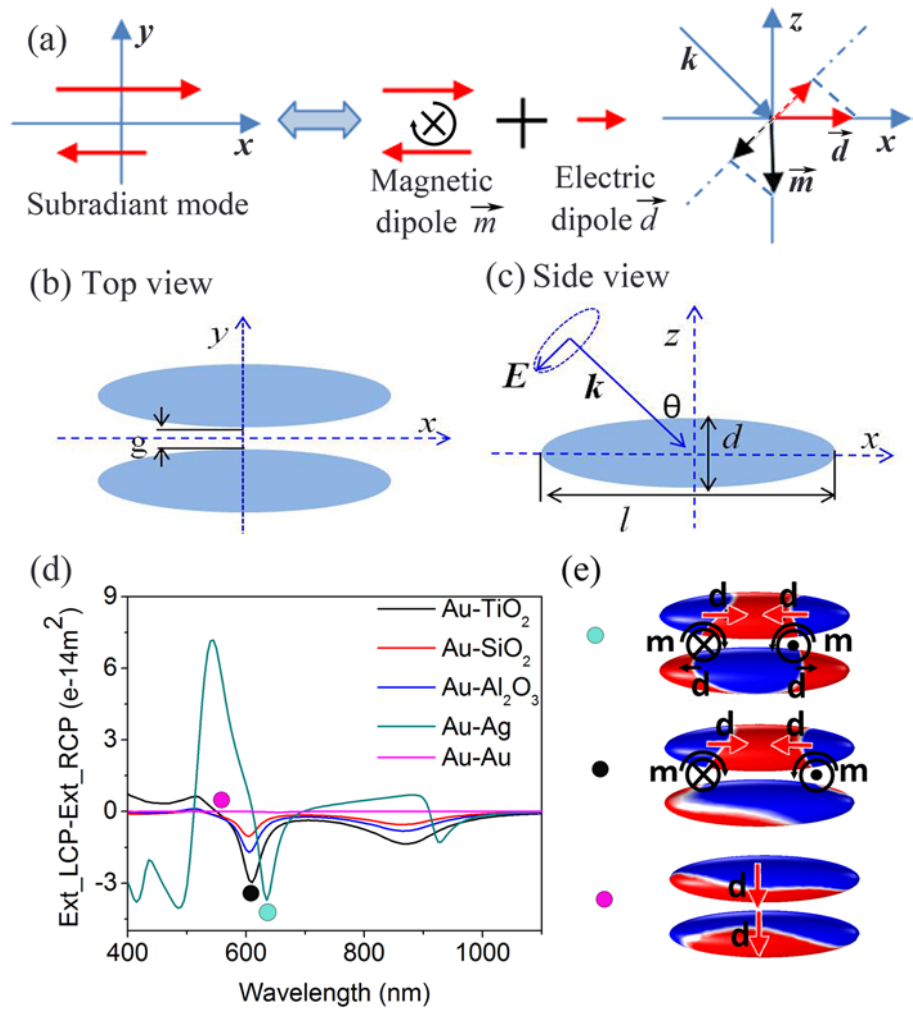

Figure 1

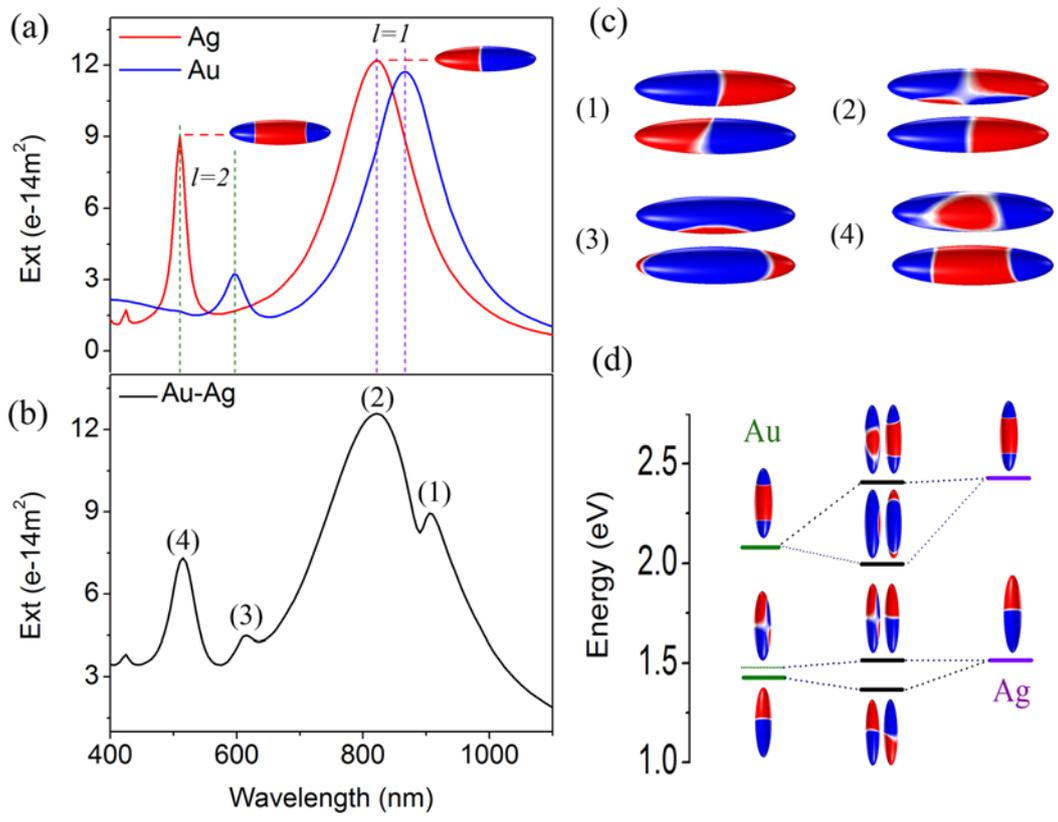

Figure 2

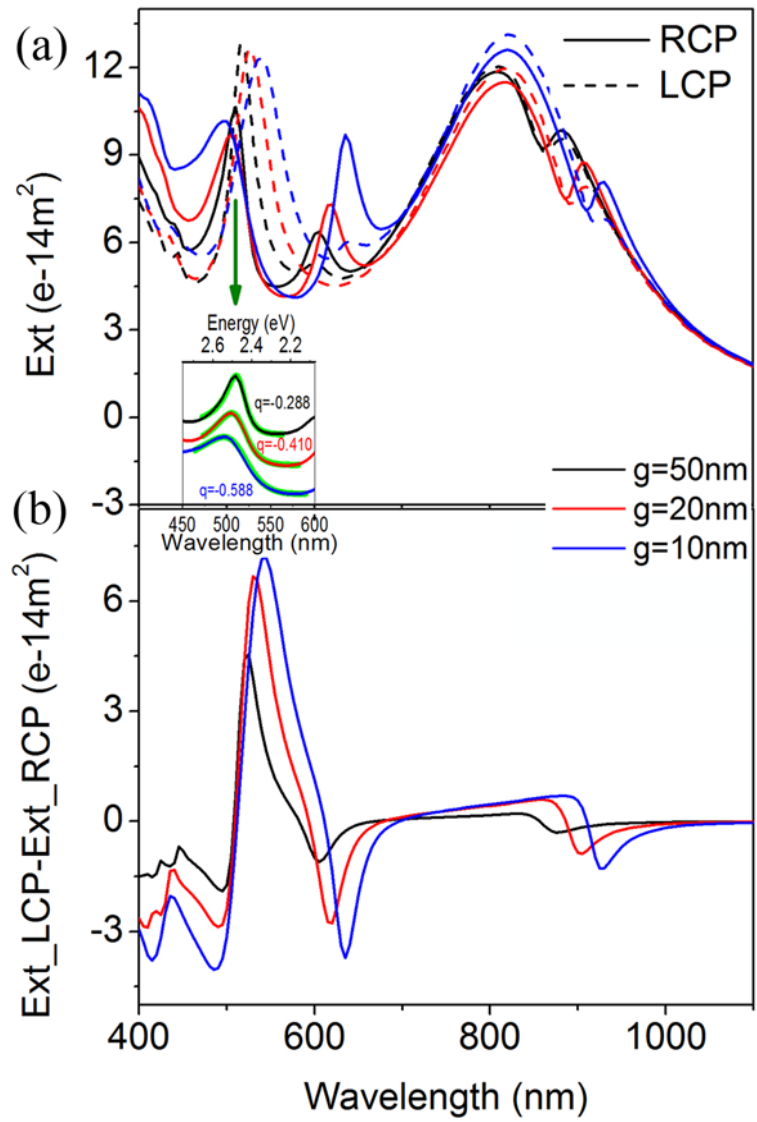

Figure 3

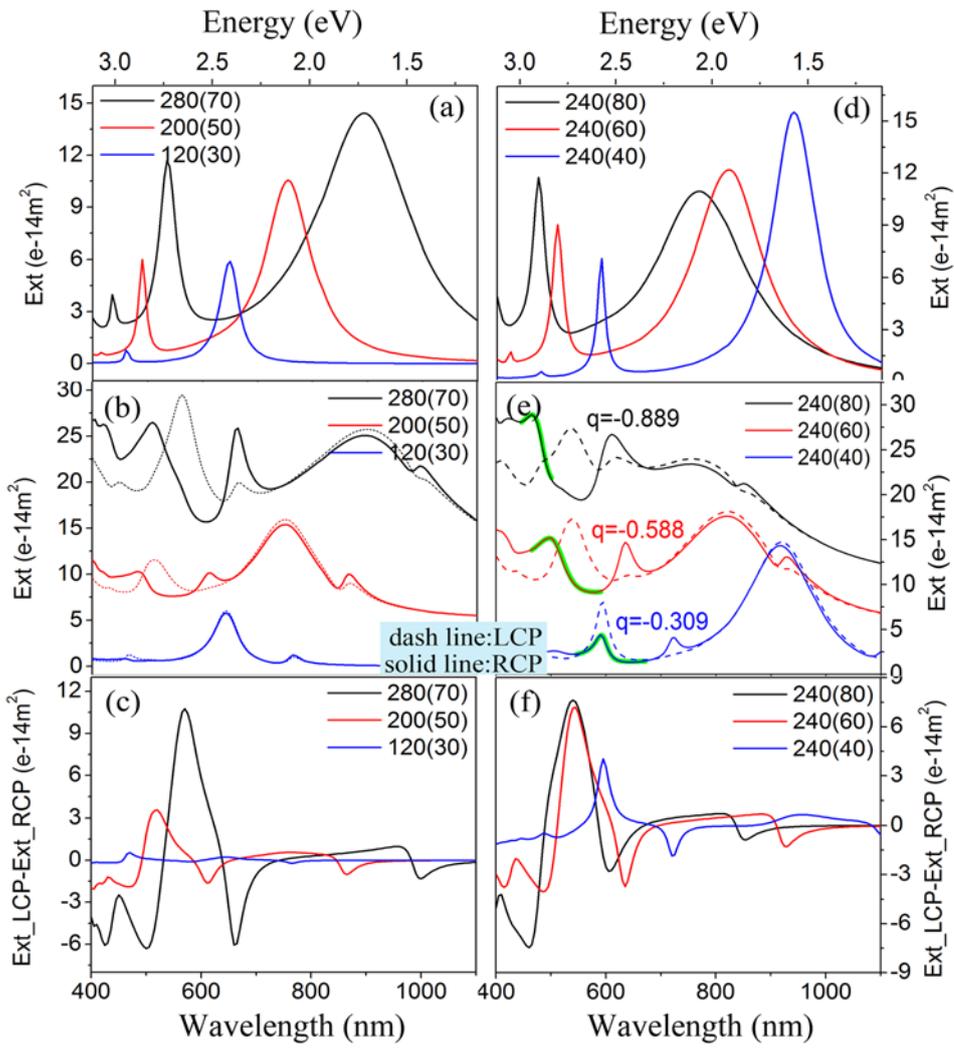

Figure 4

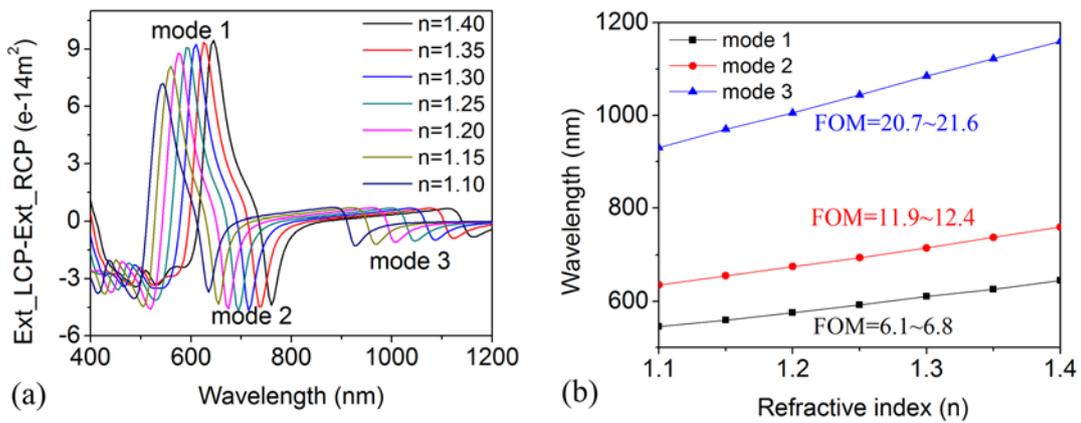

Figure 5

# Supplemental Information

# Fano resonance assisting plasmonic circular dichroism from nanorice heterodimers for extrinsic chirality

Li Hu[1,3], Yingzhou Huang[1,]*, Liang Fang[1], Guo Chen[1], Hua Wei[1], Yurui Fang[2,]*

[1] Soft Matter and Interdisciplinary Research Center, College of Physics, Chongqing University, Chongqing, 400044, P. R. China

[2] Department of Applied Physics, Chalmers University of Technology, Göteborg, SE 412 96, Sweden

[3] School of Computer Science and Information Engineering, Chongqing Technology and Business University, Chongqing, 400067, China

Corresponding Authors. Emails: yzhuang@cqu.edu.cn (Y. Huang), yurui.fang@chalmers.se (Y. Fang)

**The influence of materials on plasmon resonance.** Considering the two nanorices in each dimer were of the same size here, the materials of the two nanorices should be different. Therefore, various groups of nanorice materials were investigated. Figure S1 shows the extinction spectra of the heterodimer (l=240 nm, d=60 nm, g=10 nm, n=1.1) with different materials illuminated by LCP and RCP light respectively. As figure S1(b) shown, when the heterodimer consists of the same material (Au-Au), the two extinction spectra are identical which excited by LCP and RCP respectively. While the heterodimers consist of Au and other materials，there are abviously different of the extinction cross section excited by LCP and RCP, as figure S1(b-e) shown.

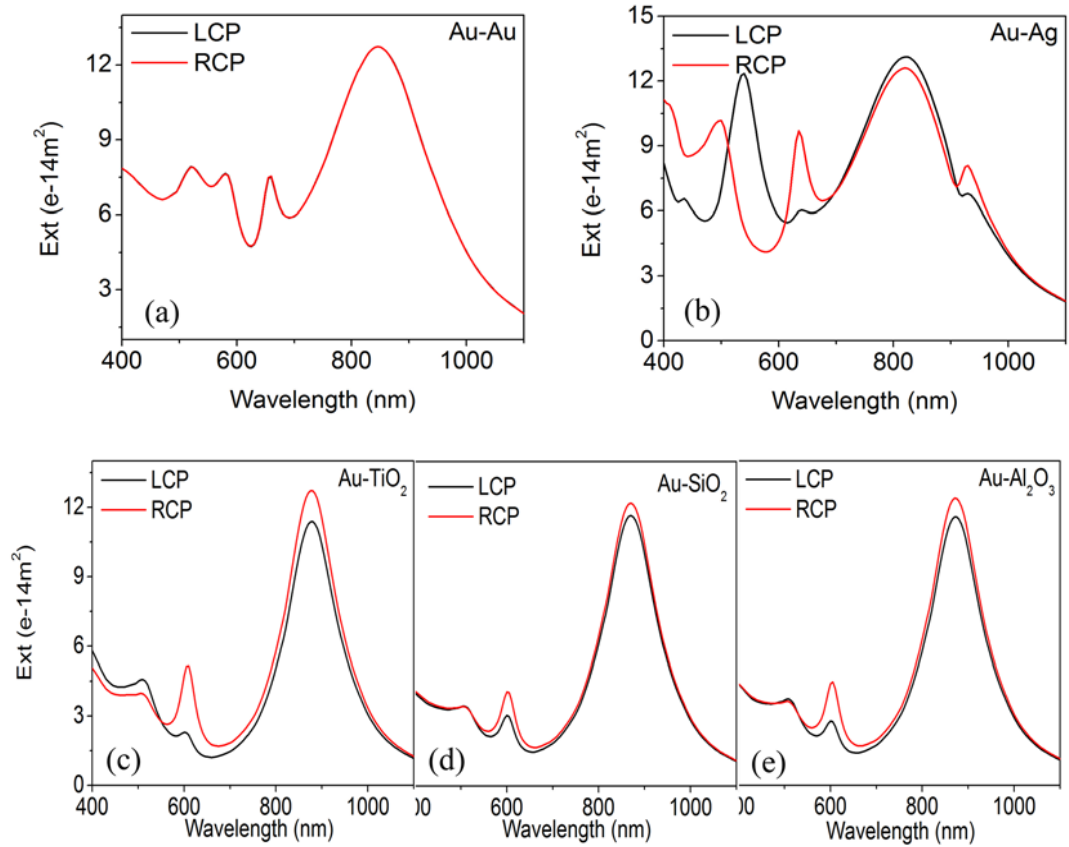

**Figure S1 | The influence of materials on plasmon resonance.** Extinction spectra of the heterodimer with different materials ($l=240$ nm, $d=60$ nm, $g=10$ nm, $\theta=\pi/4$) excited by CPL (black: LCP, red: RCP). (a) Au-Au; (b) Au-Ag; (c) Au-TiO$_2$; (d) Au-SiO$_2$; (e) Au-Al$_2$O$_3$. CPL: circularly polarized light.

**The effect of the structure factors on plasmon resonance.** The structure factors (e.g. size and shape) play a great impact on the surface plasmon and Fano resonance. Figure S2(a) show the extinction cross section of Au and Ag nanorice with various sizes but the same aspect ratio ($l/d= 4$) excited by linear polarized light. As Figure S2(a) shown, the intensity of the resonant modes especially the quadruple mode increases markedly with the size of Ag and Au nanorice increasing, and the resonant wavelength have a significant red-shift . At the same time, the distance increases between the dipole and quadruple modes. Figure S2(b) shows the extinction spectra of the Au-Ag heterodimer excited by circularly polarized light (CPL). Fig S2(c) shows the extinction cross section of Au and Ag nanorice with various aspect ratio ($l/d = 3, 4, 6$) excited by linear polarized light. It can be seen the variation trend of Au nanorice and Ag nanorice is similar, and the Au nanorice change smaller. Figure S2(d) shows the extinction spectra of the heterodimer excited by LCP and RCP light .

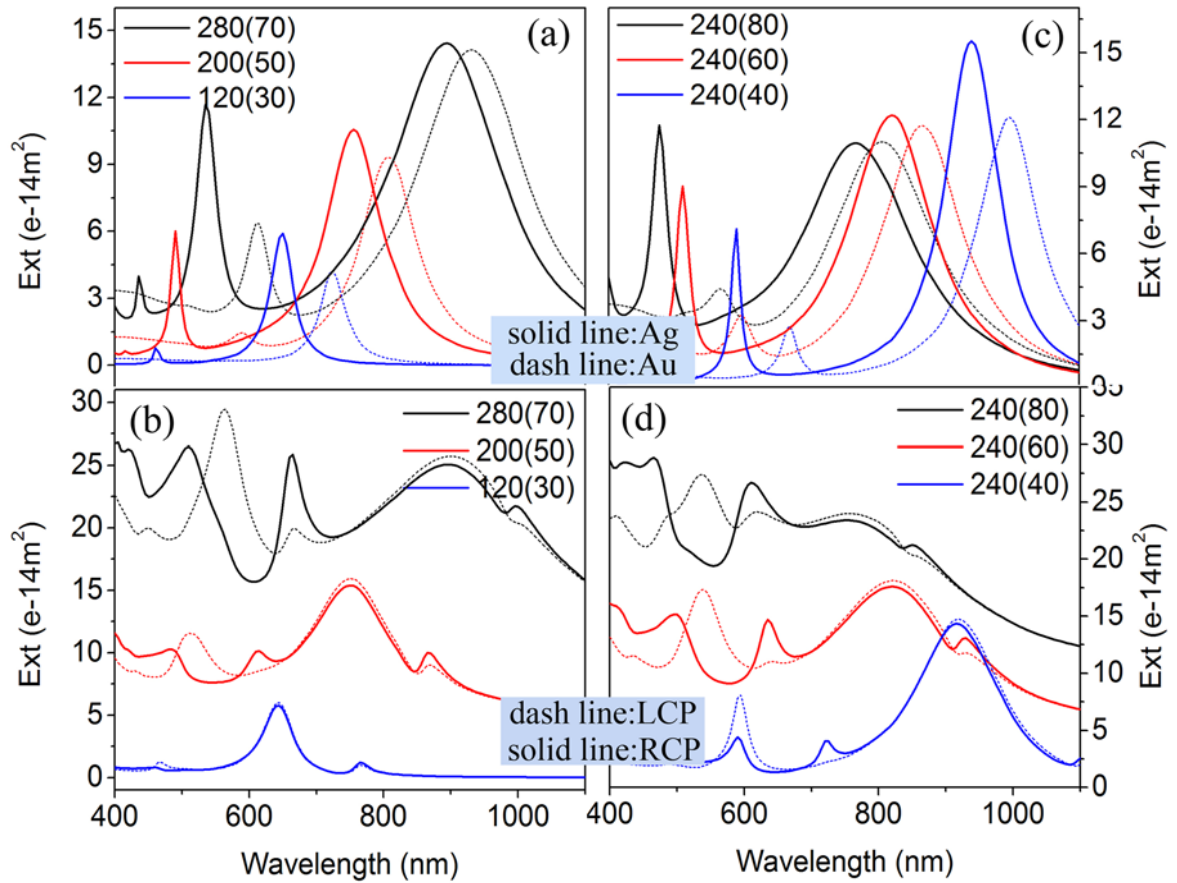

**Figure S2 | The effect of the structure factors on plasmon resonance..** (a) The extinction spectra of individual Ag nanorice and Au nanorice excited by linear polarized light ($l/d=4$, $g=10$ nm). (b) The extinction spectra of Au-Ag heterodimer excited by LCP and RCP light ($l/d=4$, $g=10$ nm). (c) The extinction spectra of single Ag nanorice and Au nanorice excited by linear polarized light ($l/d=3,4,6$, $g=10$ nm). (d) The extinction spectra of Au-Ag heterodimer excited by LCP and RCP light. Set a heterodimer as ($l(d)$) in figure.

**Sensitivity of the Fano resonances to the surroundings.** Fig S3 (a) shows the extinction spectra of Au-Ag heterodimer in different surroundings excited by linearly polarized light. We can see that there are two obvious peak relatively. Fig S3 (b) shows the linear plot of resonance peak shifts of different order modes as a function of refractive index n. The *FOM* of mode 1 from 6.6 to 10.1, the *FOM* from 4.2 to 4.3 of the mode 2. Fig S4 (a) and (c) show the extinction spectra of Au-Ag heterodimer in different surroundings excited by LCP and RCP. Fig S4 (b) and (d) show the linear plot of resonance peak shifts of different order modes as a function of refractive index

n. The *FOM* of mode L1 from 5.8 to 6.2 and mode R1 from 15.6-16.1, the *FOM* from 4.8 to 5.0 for the mode L2 and R2.

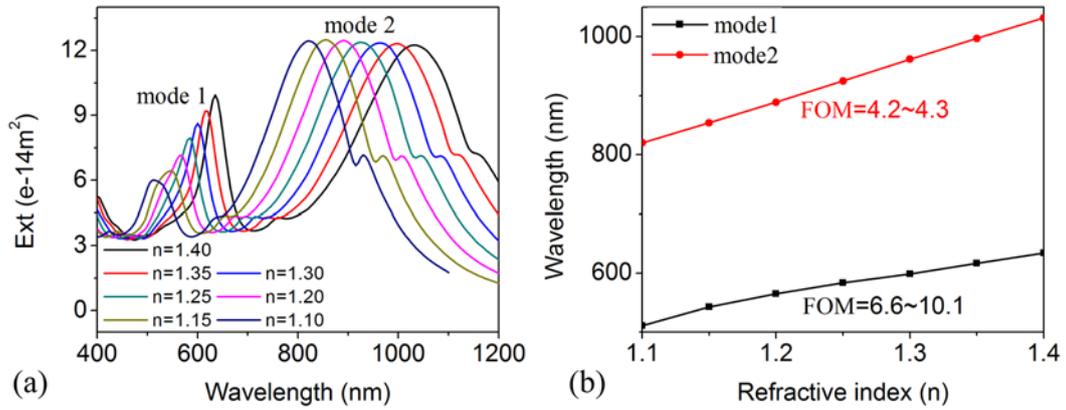

**Figure S3 | Sensitivity of the Fano resonances to the surroundings.** (a) Extinction spectra of Au-Ag heterodimer ($l = 240nm, d = 60nm, g = 10nm$) in different surroundings excited by linearly polarized light. (b) Linear plot of resonance peak shifts of different order modes as a function of refractive index n. *FOM* of each mode is calculated.

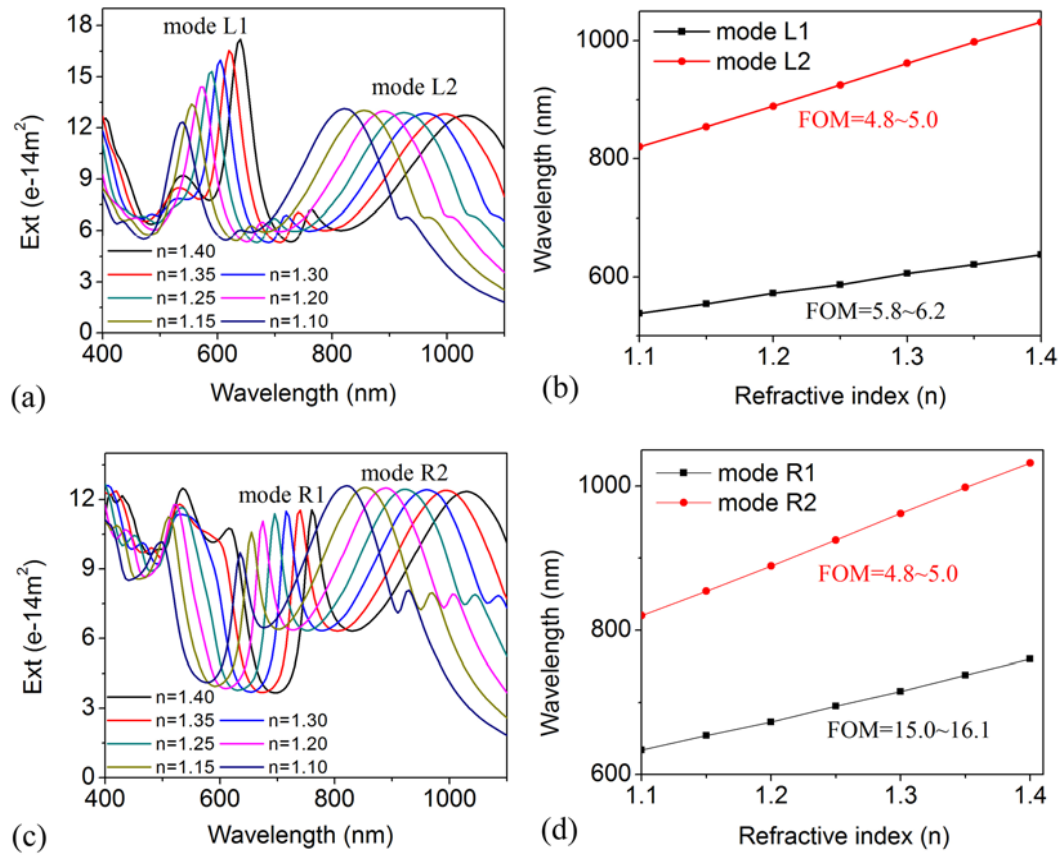

**Figure S4 | Sensitivity of the Fano resonances to the surroundings.** (a, c) Extinction spectra of Au-Ag

heterodimer ($l = 240nm, d = 60nm, g = 10nm$) in different surroundings excited by LCP and RCP. (b, d) Linear plot of resonance peak shifts of different order modes as a function of refractive index n (LCP and RCP, respectively). *FOM* of each mode is calculated.